\documentstyle[epsf,psfig,12pt,ncs]{article}
\newcommand {\ignore}[1]{}

\newcommand{\noi}{\noindent}
\newcommand{\bc}{\begin{center}}
\newcommand{\ec}{\end{center}}

\def\ifmath#1{\relax\ifmmode #1\else $#1$\fi}
\def\half{\ifmath{{\textstyle{1 \over 2}}}}

\def\3quarter{{\textstyle{3 \over 4}}}

\def\vs{\vskip}

\def\ra{\rightarrow}

\def\lf{\leaders\hbox to 1em{\hss.\hss}\hfill}
\def\21{$SU(2) \ot U(1)$}
\def\321{$SU(3) \ot SU(2) \ot U(1)$}
\def\L{\hbox{$\cal L$ }}
\def\gau{\hbox{gauge }}
\def\neu{\hbox{neutrino }}
\def\Eq#1{{Eq. (\ref{#1})}}

\def\fig#1{{Fig. (\ref{#1})}}

\def\VEV#1{\left\langle #1\right\rangle}

\def\lsim{\raise0.3ex\hbox{$\;<$\kern-0.75em\raise-1.1ex\hbox{$\sim\;$}}}
\def\gsim{\raise0.3ex\hbox{$\;>$\kern-0.75em\raise-1.1ex\hbox{$\sim\;$}}}
\def\half{{1\over 2}}
\def\beq{\begin{equation}}
\def\eeq{\end{equation}}
\def\bef{\begin{figure}}
\def\eef{\end{figure}}
\def\bet{\begin{table}}
\def\eet{\end{table}}
\def\bea{\begin{eqnarray}}
\def\ba{\begin{array}}
\def\ea{\end{array}}
\def\bi{\begin{itemize}}
\def\ei{\end{itemize}}
\def\ben{\begin{enumerate}}
\def\een{\end{enumerate}}
\def\ra{\rightarrow}
\def\ot{\otimes}
\def\eea{\end{eqnarray}}
\def\ib#1#2#3{           {\it ibid. }{\bf #1} (19#2) #3}

\def\nps#1#2#3{        {\it Nucl. Phys. B (Proc. Suppl.) }{\bf #1} (19#2) #3} 
\def\np#1#2#3{           {\it Nucl. Phys. }{\bf #1} (19#2) #3}
\def\pl#1#2#3{           {\it Phys. Lett. }{\bf #1} (19#2) #3}
\def\pr#1#2#3{           {\it Phys. Rev. }{\bf #1} (19#2) #3}

\def\prl#1#2#3{          {\it Phys. Rev. Lett. }{\bf #1} (19#2) #3}

\def\zp#1#2#3{           {\it Zeit. fur Physik }{\bf #1} (19#2) #3}

\def\n.c.#1#2#3{         {\it Nuovo Cim. }{\bf #1} (19#2) #3}
\def\r.n.c.#1#2#3{       {\it Riv. del Nuovo Cim. }{\bf #1} (19#2) #3}
\def\sjnp#1#2#3{         {\it Sov. J. Nucl. Phys. }{\bf #1} (19#2) #3}

\def\ppnp#1#2#3{           {\it Prog. Part. Nucl. Phys. }{\bf #1} (19#2) #3}

\begin{document}
\def\ptmm{$p\!\!\!/_T +\mu^+ \mu^-$}
\def\pt{$p\!\!\!/_T$}
\newcommand{\beqa}{\begin{eqnarray}}
\newcommand{\eeqa}{\end{eqnarray}}
\begin{titlepage}
\begin{center}
\hfill hep-ph/9612277\\
%%\hfill FTUV/96-83\\
%%\hfill IFIC/96-92\\
\vs .5cm
{\Large \bf Single-Photon Z Decays and Small Neutrino Masses}\\
\vs 1cm
{\large J. C. Rom\~ao}
\footnote{E-mail fromao@alfa.ist.utl.pt}\\
\vs 1cm
{\it Inst. Superior T\'ecnico, Dept. de F\'{\i}sica \\
Av. Rovisco Pais, 1 - 1096 Lisboa, Codex, PORTUGAL}\\
\vs 1cm
{ \large S. D. Rindani} 
\footnote {Permanent Address: Theory Group, Physical Research Laboratory,
   Navarangpura, Ahmedabad, 380 009, India} 
{ and \large J. W. F. Valle}\\
\vs 1cm
{\it Instituto de F\'{\i}sica Corpuscular - C.S.I.C.\\
Departament de F\'{\i}sica Te\`orica, Universitat de Val\`encia\\
46100 Burjassot, Val\`encia, SPAIN\\
 http://neutrinos.uv.es}\\ 
\vs 1cm
{\large \bf Abstract}
\end{center}
\vs 1cm
\begin{quotation}
\noindent
We discuss some rare Z decay signatures associated with 
extensions of the Standard Model with spontaneous lepton number
violation close to the weak scale. We show that single-photon Z 
decays such as $Z \ra \gamma H$ and $Z \ra \gamma J J$ where $H$ 
is a CP-even Higgs boson and $J$ denotes the associated CP-odd
Majoron  may occur with branching ratios accessible to LEP 
sensitivities, even though the corresponding neutrino masses 
can be very small, as required in order to explain the deficit 
of solar neutrinos.

\end{quotation}
\end{titlepage}

\section{Introduction}

There is a large variety of ways to generate naturally small 
\neu masses which do not require one to introduce a large mass 
scale \cite{fae}.
In some of these models the neutrinos acquire mass 
only through radiative corrections \cite{zee,Babu88}.
In addition to their potential in explaining present puzzles in 
neutrino physics \cite{taup95t}, such as that of solar and
atmospheric neutrinos \cite{taup95e}, such models may give 
to many new signals at high-energy accelerator experiments \cite{beyond}.

Here we consider  {\sl radiative} schemes of neutrino mass
generation. For definiteness we focus on that introduced in ref. 
\cite{Babu88} where neutrino masses are induced at the two-loop
level. For our purposes this model will be the simplest, as it 
does not contain any scalar Higgs doublet in addition to that of 
the standard model.
Following ref. \cite{ewbaryo}, we slightly generalize the model
adding a new singlet scalar boson $\sigma$ carrying two units of 
lepton number, so that this symmetry is broken spontaneously.
This leads to the existence of a physical  Goldstone boson,  
called Majoron, denoted $J$.
One feature worth-noting here is that, although the Majoron has 
very tiny couplings to matter and \gau bosons (in particular, it 
gives no contribution to the invisible $Z$ decay width), it can 
have significant couplings to the Higgs bosons.
Since the scale at which the lepton number symmetry gets broken 
in this model lies close to the weak scale, there are a variety 
of possible phenomenological implications, such as a substantial 
Higgs boson decay branching ratio into the 
the invisible channel  \cite{JoshipuraValle92}
\begin{equation}
H \rightarrow J\;+\;J
\label{JJ}
\end{equation}

In this letter we consider the signatures associated with 
the single-photon Z boson decays such as:
\bea
\label{2}
Z \ra \gamma H, \:\:
Z \ra \gamma J , \:\:
Z \ra \gamma J J
\eea
where $H$ is a CP-even Higgs boson, and $J$ denotes the associated CP-odd
Majoron. We have calculated the possible values allowed for these decay 
branching ratios within a specific model for neutrino mass proposed
in ref. \cite{ewbaryo} and which generalizes the one first proposed 
in ref. \cite{Babu88} by introducing the Majoron.
Since the Majoron $J$ is weakly coupled to the rest of the 
particles, once produced in the accelerator, it will escape
detection, leading to a missing energy signal for the Higgs 
boson  \cite{JoshipuraValle92,inv}. In the present context the 
invisible Majoron will give rise to the single-photon $Z$-decay signal
\beq
\label{Emiss}
Z \ra \gamma E\!\!\!/_T
\eeq
It is interesting to notice that single-photon events have been 
recorded at LEP which  apparently can not be ascribed to
standard model processes \cite{opal}. 

We have shown that the branching ratios for the decays $Z \ra \gamma H$
and the Higgs-mediated decay $Z \ra \gamma J J$ can reach values comparable 
with LEP sensitivities at the Z pole. It is remarkable that such sizeable 
values occur even though the associated neutrino masses are very small, 
as required in order to explain the deficit of solar neutrinos through 
the resonant conversion effect \cite{MSW}. This happens due to the fact that 
neutrino masses are induced only radiatively, at the two-loop level.
This is in sharp contrast to the conventional Majoron model formulated 
in the seesaw context, where a large scale is introduced in order to 
account for the smallness of \neu masses \cite{CMP}.

\section{The model}

We consider a modification of the model for radiative neutrino masses
first proposed in \cite{Babu88} to incorporate spontaneous breaking of
global lepton number, leading to a majoron.

The model is based on the gauge group $SU(2)\times U(1)$, with the same
fermion content as that of the standard model, but three complex singlets 
of scalars in addition to the doublet. Thus the quark sector is standard
and no right-handed neutrino is introduced. Of the three complex singlets, 
two are charged, viz., 
$h^{\pm}$ with charge $\pm 1$ and lepton number $\mp 2$, and 
$k^{\pm \pm}$ with charge $\pm 2$ and lepton number $\mp 2$. 
The third neutral singlet scalar $\sigma$ carries lepton number 2 and 
is introduced so as to conserve lepton number in the full Lagrangian,
including the scalar self-interactions \cite{ewbaryo}. 

\begin{figure}
\centerline{\protect\hbox{\psfig{file=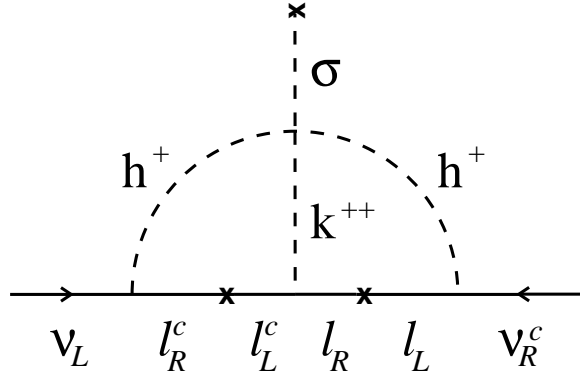,height=5cm}}}
\caption{Two-loop-induced Neutrino Mass. }
\label{2loop}
\end{figure}

With the choice of scalars and the representations which we have made,
the most general Yukawa interactions of the leptons can be written as
\beq
{\cal L} = -\frac{\sqrt2 m_i}{v}{\bar{\ell}}_i \phi e_{Ri} +
f_{ij} \ell_i^T C i \tau_2 \ell_j h^+ +
h_{ij} e_{Ri}^T C e_{Rj} k^{++} + H.c.
\label{yuk}
\eeq
where $h$ and $f$ are symmetric and anti-symmetric coupling matrices, 
respectively. The lepton masses are generated when the \21 \gau symmetry 
is broken by $\VEV{\phi}$. The first term gives the charged lepton masses 
$m_i$ at the tree level, while neutrinos acquire masses radiatively,
at the two-loop level, by the diagram in Fig. \ref{2loop}.
For reasonable and natural choices of parameters, consistent 
with all present observations, e.g. $f_{e\tau},  f_{\mu\tau}, 
h_{\tau\tau} \sim 0.01$, the singlet vacuum expectation value 
of about 100 GeV, and charged Higgs  boson masses of about 100 
GeV, these \neu masses are in the $10^{-2}$ to $10^{-3}$ eV 
range, where they could explain the deficit of solar neutrinos
through the resonant conversion effect \cite{MSW}.

\section{The scalar potential}

The most general scalar potential which is invariant under the gauge group
and under global lepton number, with at most quartic terms, is
\begin{eqnarray}
V(\phi , h, k, \sigma ) & = & \mu_1^2 \phi ^{\dagger} \phi + 
\mu_2^2 h^+ h^-	+ \mu_3^2 k^{++} k^{--} + \mu_0^2 \sigma^* \sigma \cr
&&\cr
&& + \lambda_1 (\phi ^{\dagger} \phi)^2 + \lambda_2^2 (h^+h^-)^2 + \lambda_3
(k^{++} k^{--})^2 + \lambda_0 (\sigma^* \sigma)^2 \cr
&&\cr
&& + \lambda_4 (\phi^{\dagger}\phi)(h^+h^-) + \lambda_5 (\phi^{\dagger}\phi)
(k^{++} k^{--}) + \lambda_6 (h^+ h^-)(k^{++} k^{--}) \cr
&&\cr
&& + \lambda_7 (\phi^{\dagger}\phi)(\sigma^*\sigma) 
+ \lambda_8 (h^+ h^-)(\sigma^*
\sigma) + \lambda_9 (k^{++}k^{--})(\sigma^*\sigma) \cr
\cr
&& + \lambda_0 h^+h^+k^{--}\sigma + \lambda_0^* h^-h^-k^{++}\sigma^*.
\label{pot}
\end{eqnarray}

We assume that for a choice of parameters of the potential, 
both the \21 gauge symmetry as well as the global lepton number 
symmetry are broken spontaneously, with the neutral scalar fields
getting vacuum expectations values. We rewrite the
neutral fields as follows:
\beq
\phi^0 = \frac{1}{\sqrt{2}} ( v + \phi_R^0 + i \phi_I^0),
\label{phi}
\eeq
and
\beq
\sigma = \frac{1}{\sqrt{2}} ( \omega + \sigma_R  + i \sigma_I).
\label{sig}
\eeq
$v$ and $\omega$ are the vacuum expectation values defined by
\footnote{Our choice of the $\phi$ vacuum expectation value differs 
from that in \cite{Babu88} by a factor of $\sqrt{2}$}
\beq
\langle \phi^0 \rangle = \frac{v}{\sqrt{2}},
\eeq
\beq
\langle \sigma \rangle = \frac{\omega}{\sqrt{2}}.
\eeq

The physical massive
scalars which survive are those corresponding to $h^{\pm}$, $k^{\pm \pm}$, 
and two orthogonal combinations of 
of $\phi_R^0$ and $\sigma_R$. The charged components of $\phi$, viz.,
$\phi^{\pm}$, correspond to the would-be Goldstone particles absorbed
by $W^{\pm}$, $\phi^{0}_I $ is the would-be Goldstone eaten by the
$Z$ boson and $\sigma_I$ is the massless physical Goldstone field 
corresponding to spontaneously broken global lepton number.

We can write the following expressions for the squared masses of the 
various scalars: 
\beq
M_{h^+}^2 = \mu_2^2 + \half \lambda_4 v^2 + \half \lambda_8 \omega ^2,
\eeq
\beq
M_{k^{++}}^2 = \mu_3^2 + \half \lambda_5 v^2 + \half \lambda_9 \omega ^2.
\eeq
The squared mass terms for the neutral scalars can be written as 
$
 - \half \Phi_i M^2_{ij} \Phi_j + \cdots
$ 
where we have defined the vector
\beq
\Phi=\left[\matrix{
\phi^0_R \cr
\sigma_R \cr}
\right].
\eeq
The squared mass matrix $M^2$ is given by
\beq
M^2 = \left[ \matrix{
2 \lambda_1\ v^2 & \lambda_7\ \omega v \cr
\lambda_7\ \omega v & 2 \lambda_0\ \omega^2 \cr}
\right]
\eeq
The mass eigenstates are $H_i$ defined through
\beq
H_i=P_{ij} \Phi_j
\label{eigen}
\eeq
where the diagonalization matrix $P$ is orthogonal, that is, $P^{-1}=P^T$.
Therefore the inverse of \Eq{eigen} reads
\beq
\Phi_i= P_{ji}\ H_j
\eeq
or in terms of the fields $\phi^0_R$ and $\sigma_R$
\beq
\left\{
\begin{array}{l}
\phi^0_R= P_{11}\ H_1 + P_{21}\ H_2 \cr
\sigma_R= P_{12}\ H_1 + P_{22}\ H_2 \cr
\end{array}
\right.
\label{rel2}
\eeq

Before we close this section let us derive two important relations. In the
basis $\Phi_i$ the eigenvectors $H_i$ have components
\beq
\left[
\matrix{
P_{i1} \cr
P_{i2} \cr}
\right]
\eeq
Therefore the eigenvalue equation reads
\beq
M^2 \ H_i = M^2_{H_i}\ H_i \quad , \quad i=1,2
\eeq
which gives explicitly
\beq
\begin{array}{l}
2 \lambda_1\ v^2\ P_{i1} + \lambda_7\ \omega v\ P_{i2}=M^2_{H_i}\ P_{i1} \cr
\lambda_7\ \omega v\ P_{i1} + 2 \lambda_0\ \omega^2\ P_{i2}= M^2_{H_i}\
P_{i2} 
\end{array}
\label{rel1}
\eeq
These expressions will be useful below. 

\section{The calculation of the single-photon processes}

In this section we will describe the relevant couplings which are
different from the ones in standard model, or are new. 
The couplings of the physical and unphysical scalars among themselves 
are simply obtained by substituting from \Eq{phi} and \Eq{sig} into 
the scalar potential given by \Eq{pot}, and making use of \Eq{rel1} 
and \Eq{rel2}. The relevant terms in the Lagrangian resulting from
this substitution are:
\beqa
-\L&=&
 \phi^+\phi^-H_i\ \frac{M^2_{H_i}}{v}\ P_{i1} 
+ \half J^2 H_i\ \frac{M^2_{H_i}}{\omega}\ P_{i2}\cr
&& + \half J^2 \left[ \lambda_7\  \phi^+\phi^- 
+ \lambda_8\  h^+h^- 
+ \lambda_9\  k^{++}k^{--} \right] + \cdots .
\eeqa

The unphysical scalars $\phi^{\pm}$ have exactly the same couplings to the 
gauge fields and the Faddeev-Popov ghosts as in the standard model,
whereas the couplings of the neutral massive scalars, $H_i$, are obtained 
by multiplying the standard model couplings, written in terms of the 
physical masses, by $P_{i1}$. For example, the coupling of
$H_i$ to $W^+W^-$ is given by
\beq
{\cal L}= g M_W P_{i1}\ W^+_{\mu} W^{- \mu} H_i.
\eeq

The charged physical scalars $h$ and $k$ have the following couplings to 
the gauge bosons:
\beqa
\L& =& - ie(A_{\mu} +  \tan\theta_W Z_{\mu}) \left\{ (h^- \partial_{\mu} h^+
-  \partial_{\mu} h^- h^+) \right. \cr 
&& \cr
&& \left. + 2(k^{--} \partial_{\mu} k^{++}  
-  \partial_{\mu}k^{--} k^{++})  \right\} \cr 
&&\cr
&& + e^2 (A_{\mu} + \tan\theta_W Z_{\mu})^2 (h^+h^- + 4 k^{++}k^{--}).
\eeqa

In order estimate the branching ratios for the single-photon
processes in our model we have varied the values of $M_{H_2}$, 
of $M_{h^{\pm}}$, of $M_{k^{\pm\pm}}$ in the 100 GeV range,
and the quartic couplings in the potential over the range
\bea
0 \leq & \lambda_{\mbox{quartic}} & \leq \sqrt{4\pi}
\eea
while the lepton number violation scale ${\omega}$ and CP-even Higgs
mixing angle $\theta$  were chosen in the range
\beqa
2 \leq &\displaystyle \frac{v}{\omega}& \leq 3 \cr
&&\cr
0 \leq &\theta& \leq \frac{\pi}{2} 
\eeqa
\noi
We have also studied the effect having lower values for the lepton 
number violation scale ${\omega}$, and obtained a slight enhancement 
of our branching ratios for the single-photon processes. Notice that with 
our conventions we have for the mixing matrix of the CP-even Higgs bosons
\beq
P=\left[\ba{lr}
\cos \theta & -\sin \theta \cr
\sin \theta & \cos \theta 
\ea
\right]
\eeq

\subsection{ The $Z \rightarrow H \gamma $ decay}
\label{zhg}

\begin{figure}
\centerline{\protect\hbox{\psfig{file=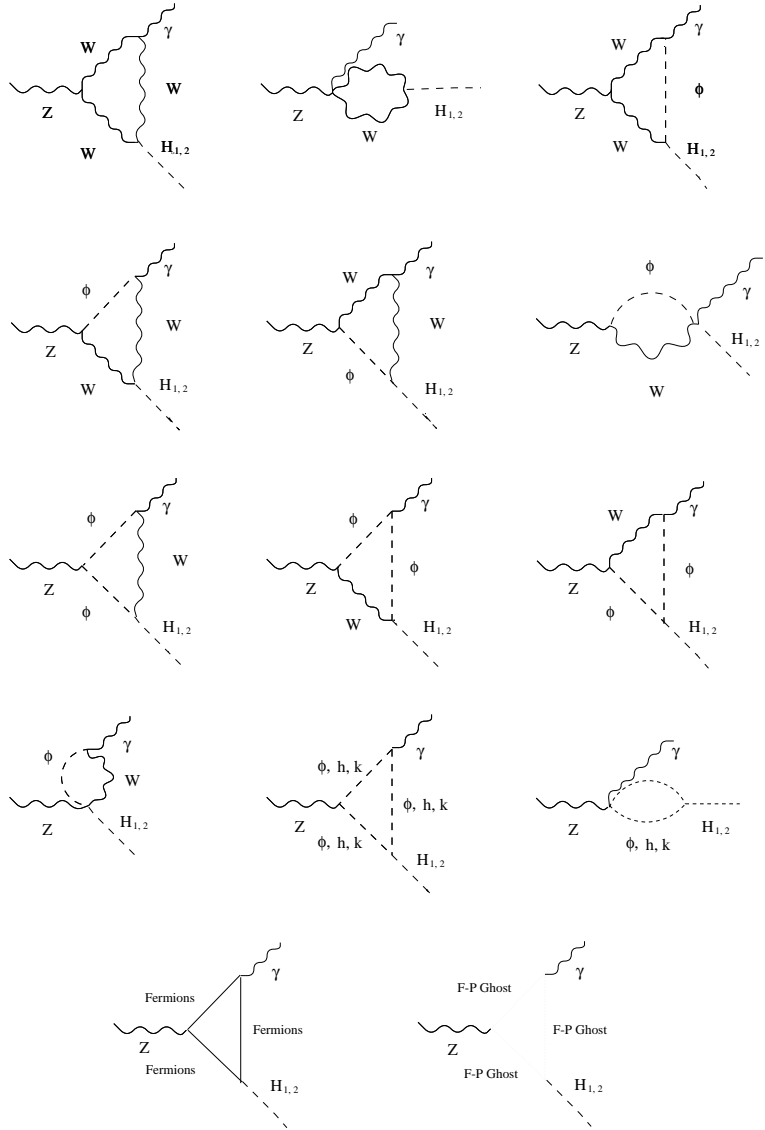,height=15cm}}}
\caption{Feynman diagrams for the decay $Z \rightarrow H \gamma$. }
\label{diagDnew}
\end{figure}

This process arises from the Feynman diagrams shown in \fig{diagDnew}. 
In addition to standard model diagrams this process receives 
contributions from the new physical singly as well doubly charged 
scalar bosons, as shown in \fig{diagDnew}. The amplitude for the 
process can be written as
\beq
{\cal M}= \epsilon_Z^{\mu} \epsilon_A^{\nu}\ \frac{e g^2}{16\pi^2 M_W}\
\left( g_{\mu \nu} q_1 \cdot q_2 - q_{1\mu} q_{2\nu} \right)\
A_{H \gamma}
\eeq
where $q_1$ and $q_2$ are the photon and Higgs momenta,
respectively. The normalized amplitude $A_{H \gamma}$ is given by
\beq
A_{H \gamma} = A_{SM}\ P_{11} + A_h + A_k
\eeq
where $A_{SM}$ is the corresponding amplitude 
for the standard model and $A_h$ and $A_k$ are the amplitudes 
corresponding to the loops of the new charged scalars. We give their
explicit expressions in the Appendix.

The resulting $Z \ra \gamma H$ decay width is then
\beq
\Gamma=\frac{1}{12\pi}\ \left( \frac{e g^2}{16 \pi^2 M_W} \right)^2\
E_{\gamma}^3\ |A_{H \gamma} |^2
\eeq
where $E_{\gamma}=(M_Z^2-M_h^2)/(2M_Z)$ is the energy of the photon.

\begin{figure}
\centerline{\protect\hbox{\psfig{file=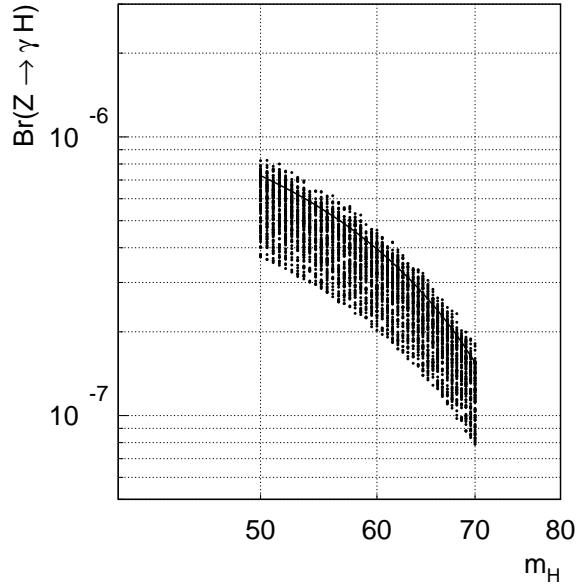,height=8cm}}}
\caption{Branching ratio for $Z \ra \gamma H$ as a function of $M_H$ 
for the standard model (solid line) and for our model (points). }
\label{babu1_10070}
\end{figure}

As an illustrative example we show in \fig{babu1_10070} the expected
branching ratio branching ratio for $Z \ra \gamma H$ as a function 
of $M_H$ for the standard model and for our model
\footnote{In the present model we also denote $M_H$ the mass of
the lightest CP-even Higgs boson $H_1$.}. 
In this Figure we have taken $M_{H_2}=100\ GeV$ and 
$M_{h^{\pm}}=M_{k^{\pm\pm}}=70\ GeV$. For reasonable allowed
choices of the relevant parameters one sees that this the value 
of this branching ratio can be enhanced with respect to 
the standard model predictions,  but only slightly, by
a factor 2 or so, for any fixed  $M_H$.
The most novel aspect of this decay in the present model comes 
from the fact that CP-even Higgs boson $H$ may decay into two 
Majorons with a substantial branching ratio, leading to a mono-photon 
plus missing energy signature for the decay $Z \rightarrow H \gamma$.

\subsection{ Majoron emitting $Z$ decays}

The majoron does not couple to the Z boson at the tree level,
since it is an \21 singlet. Nevertheless it can couple
radiatively leading, for example, to  processes such as 
$Z \rightarrow \gamma J$ and $Z \rightarrow \gamma + J + J$, 
recently discussed in a different context in ref. \cite{mono}. 
These processes are, of course, absent in the standard model.

The single majoron emission process would give rise to a
characteristic signature consisting of monochromatic photons
plus missing energy. In contrast to the model considered in ref. 
\cite{mono}, the single majoron emission process is expected 
to be very small in the present model. Notice, for example, 
that since the majoron does not couple to charged leptons 
at the tree level, the one-loop diagram involving charged 
lepton exchange is absent.

%%\subsection{ $Z \rightarrow \gamma JJ$}

\begin{figure}
\centerline{\protect\hbox{\psfig{file=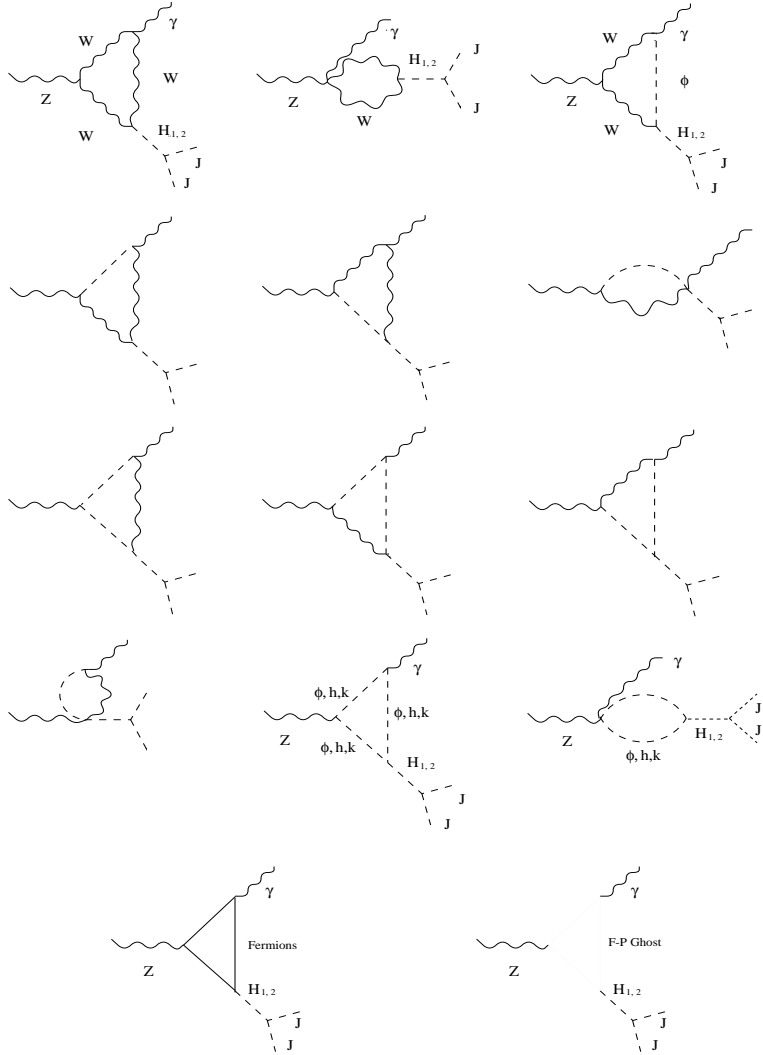,width=10cm,height=14cm}}}
\caption{Feynman diagrams for the $H$-mediated decay 
$Z \rightarrow \gamma + J + J$ }
\label{diagB}
\end{figure}

\begin{figure}
\centerline{\protect\hbox{\psfig{file=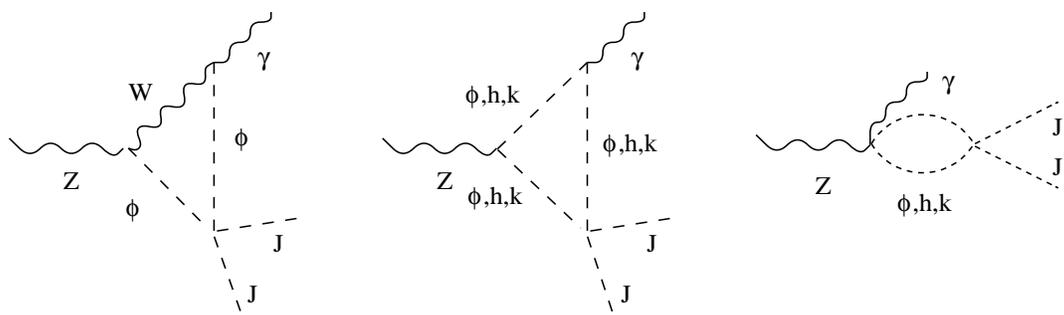,width=14cm}}}
%\vglue -13cm
\caption{Additional Feynman diagrams for the decay $Z \ra \gamma + J + J$. }
\label{diagC}
\end{figure}

In contrast the process $Z \ra \gamma JJ$ proceeds at the one-loop level
 through two types of diagrams. The first  set of diagrams involves 
the one-loop coupling of $Z$ to $\gamma$ and $H_i$ (which may be off-shell), 
with a tree-level coupling of $H_i$ to two majorons, \fig{diagB}. 
In the other set of diagrams \fig{diagC}  
the two majorons arise from a quartic coupling to 
a pair of charged scalars ($\phi^{\pm}$, $h^{\pm}$ or $k^{\pm\pm}$). The first 
set is directly related to the set of diagrams for the process $Z \ra
\gamma H_1$ discussed in Section \ref{zhg}, and can 
be computed simply by first replacing the Higgs boson $H_1$ by 
$H_i$, then multiplying the corresponding amplitude by 
the propagator for the field $H_i$, and finally summing over $i=1,2$. 
The particles running in the loops are, in this case, not only those 
present in the standard model but also the charged
scalars $h^{\pm}$ and $k^{\pm\pm}$. The corresponding amplitudes have
been calculated in Section \ref{zhg}.  
The second set of diagrams correspond to the quartic couplings of the
majorons to the charged scalars. It can be shown that the first
diagram of \fig{diagC} is not gauge invariant but exactly cancels
against the non gauge invariant part of the diagrams in \fig{diagB}. 
The remaining diagrams of \fig{diagC} with $\phi^{\pm}$, $h^{\pm}$ and 
$k^{\pm\pm}$ running in the loop are gauge invariant by themselves and 
have to be calculated afresh.

\noi
Gauge invariance and $CP$ conservation allow us to write the amplitude for 
\beq
Z(P) \rightarrow \gamma(q_1) + J(q_2) + J(q_3)
\label{pro}
\eeq
for the case of on-shell $Z$ and $\gamma$, as 
\beq
M=\epsilon^Z_{\mu}\, \epsilon^{\gamma}_{\nu}\, 
\frac{eg^2}{16 \pi^2 M_W}\, (g^{\mu \nu} q_1 \cdot Q 
- q_1^{\mu} Q^{\nu})\, A_{\gamma J J} 
\eeq
where $Q=q_2+q_3$ and use has been made of current conservation 
for on-shell $Z$ and $\gamma$.

\noi
The combined contribution of the first set of diagrams to 
$A_{\gamma J J}$ can be deduced from the result of Section \ref{zhg}. 
The answer is:
\beq
A_{\gamma J J}^{(1)}=\sum_{i=1}^2\ 
\frac{A_{H \gamma}(Q^2)}{Q^2-M^2_{H_i}+i M_{H_i} \Gamma_{H_i}}\,
\frac{M^2_{H_i}}{\omega}\ P_{i2}
\label{azgjj1}
\eeq
where $A_{H \gamma}(Q^2)$ is the amplitude calculated in
Section \ref{zhg} evaluated at $Q^2=(P-q_1)^2=M_Z(M_Z-2E_{\gamma})$.
In \Eq{azgjj1} we have introduced the width of $H_i$, because, as we
shall see, the dominant contribution for the process comes when
$Q^2\simeq M^2_{H_i}$. 

\noi
The contribution to $A_{\gamma J J}$ of the second set of diagrams can
be written as
\beq
A_{\gamma J J}^{(2)} = \hat{A}_{\phi} +\hat{A}_h +\hat{A}_k
\eeq
where $\hat{A}_{\phi}$, $\hat{A}_h$ and $\hat{A}_k$ are the
contributions of the charged scalars (unphysical and physical) and are
given explicitly in the Appendix.

\noi
The photon energy spectrum is then
\beq
\frac{d \Gamma}{d E_{\gamma}}=\frac{1}{192\, \pi^3}\, 
\left( \frac{e\, g^2}{16\, \pi^2\, M_W}\right)^2\
M_Z\, E^3_{\gamma}\ \left| A_{\gamma J J}^{(1)}+A_{\gamma J J}^{(2)}\right|^2
\eeq
and the total width
\beq
\Gamma= \int_0^{\half M_Z} \frac{d \Gamma}{d E_{\gamma}}
{d E_{\gamma}}
\eeq

\noi
We have explicitly verified that the contribution of $A_{\gamma J J}^{(2)}$ 
is  small when compared with the standard model result for 
$Z\ra H \gamma$. Thus the main contribution comes from the first 
set of diagrams when $Q^2 \simeq M_{H_i}$. For this reason we need 
to evaluate the width of $H_i$. As an approximation, we assume that 
the doublet part of $H_i$ decays mainly in $\overline{b}b$. 
In this case we have only two partial widths
\beq
\Gamma(H_i \ra JJ)= \frac{1}{32\, \pi}\ \frac{g^2_{H_i JJ}}{M_{H_i}}
\eeq
and
\beq
\Gamma(H_i\ra \overline{b}b)= \frac{1}{4\, \pi}\ M_{H_i}\ 
g^2_{H_i \overline{b} b}\ \left(1- \frac{4\, m^2_b}{M^2_{H_i}} \right)^{3/2}
\eeq
where
\beq
g_{H_i JJ}= \frac{M^2_{H_i}}{\omega}\ P_{i2}
\quad \hbox{and} \quad 
g_{H_i \overline{b} b}= \frac{m_b\,P_{i1}}{v}
\label{hcoup}
\eeq
As we said before the photon energy spectrum is peaked around 
$E_{\gamma}=(M^2_Z-M^2_{H_i})/(2M_Z)$. However this does not mean that
the contribution of the charged scalars is negligible. In fact, we
have two extreme cases:

%%\bs
\noi$\bullet$
{$P_{11}$ large (small $\theta$)}
%%\bs

\begin{figure}
\centerline{\protect\hbox{\psfig{file=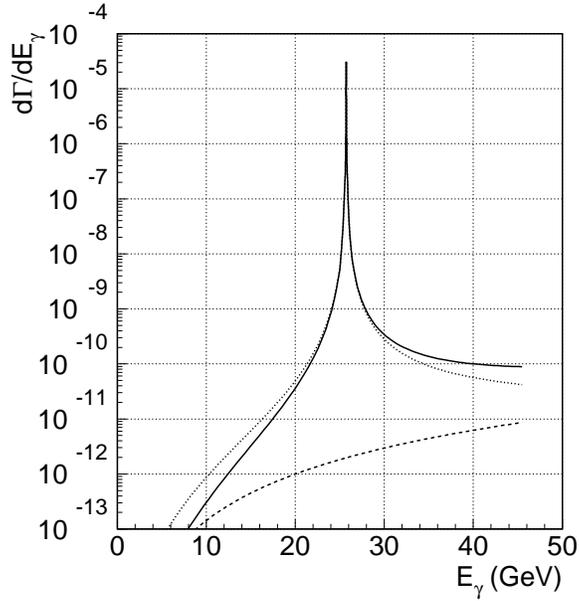,height=8cm}}}
\caption{Photon spectrum for $M_H=60$ GeV. It is peaked around
$E_{\gamma}\simeq 25.7$ GeV. The solid line represents the total
contribution, the dotted line the contribution from the resonant
diagrams, and the dashed line the contribution from the non-resonant 
ones. In this Figure we have $M_{H_2}=100\ GeV$ 
and $M_{h^{\pm}}=M_{k^{\pm\pm}}=70\ GeV$.}
\label{babu3}
\end{figure}

\begin{figure} 
\centerline{\protect\hbox{\psfig{file=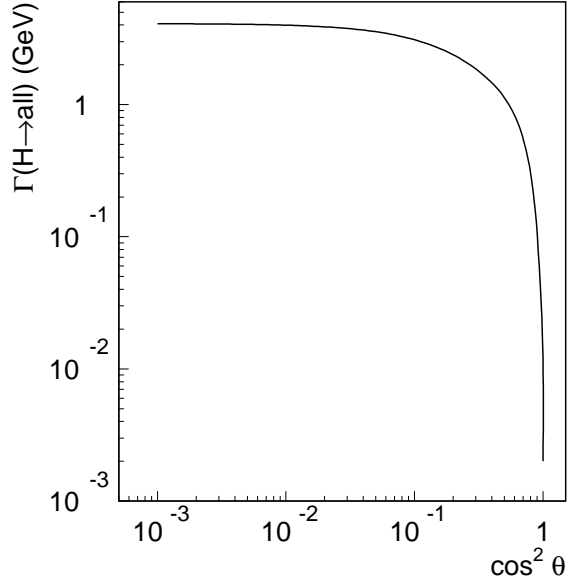,height=8cm}}}
\caption{Upper limits for  $\Gamma(H\ra all)$ as a function of 
$P_{11}=\cos^2(\theta)$.
In this Figure we have $M_{H_2}=100\ GeV$ 
and $M_{h^{\pm}}=M_{k^{\pm\pm}}=70\ GeV$.}
\label{hwidth}
\end{figure}

\noi
The dominant contribution comes from the resonant diagrams (first
set). The contribution from the loops of charged scalars with quartic vertices 
is negligible. The energy spectrum is peaked around 
$E_{\gamma}=(M_Z^2-M_H^2)/(2M_Z)$. This can be seen from \fig{babu3}
which is for $P_{11}=0.94$. Note that the other diagrams with charged
scalars are not negligible because they are also resonant. In fact it
is necessary to have them of the same order as the standard model for
$Z\ra \gamma H$ in order to have an increase.
Note also from \fig{babu3}
that the width of the $H_1$ is very small. This depends on $P_{11}$ 
being large as can be seen from \Eq{hcoup} 
and in \fig{hwidth}.

%%\bs
\noi$\bullet$
{$P_{11}$ small ($\theta$ close to $\pi/2$)}
%%\bs

\begin{figure}
\centerline{\protect\hbox{\psfig{file=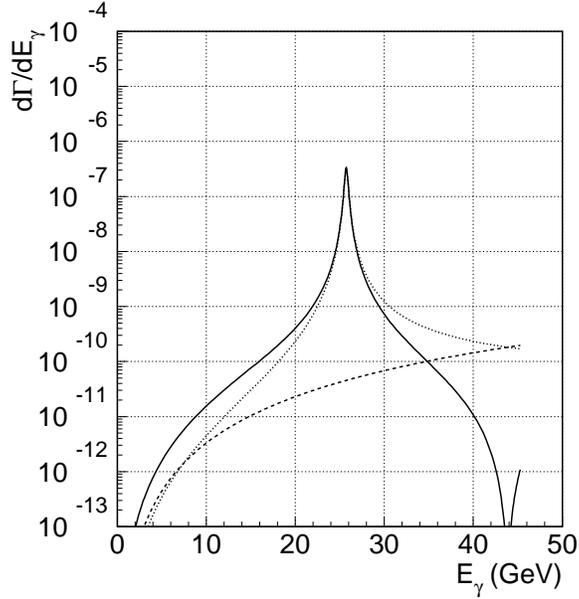,height=8cm}}}
\caption{Photon spectrum for $M_H=60$ GeV. It is peaked around
$E_{\gamma}\simeq 25.7$ GeV. The solid line represents the total
contribution, the dotted line the contribution from the resonant
diagrams, and the dashed line the contribution from the non resonant ones.
In this Figure we have $M_{H_2}=100\ GeV$ 
and $M_{h^{\pm}}=M_{k^{\pm\pm}}=70\ GeV$.}
\label{babu3a}
\end{figure}

\noi
Now the standard model-like diagrams are small and the main contribution 
is from the loops of charged scalars. However the main contribution is still
from the resonant charged scalars diagrams. The non-resonant diagrams
are small, although not completely negligible. In \fig{babu3a} we
illustrate this for $P_{11}=0.04$. 
There we can also see that the width of the $H_1$ is a few GeV's in
agreement with \fig{hwidth}.
Note that when $\theta \sim \pi/2$ the branching ratio of the Higgs to
JJ is close to one. Thus the standard way of looking for the Higgs, 
through the standard $b-\bar{b}$ decay mode, would miss it. In the
present context this implies that, in addition to the broad photon 
spectrum in the photon + missing energy signal, one has the additional 
feature that the $\gamma + b\bar{b}$ signal would be weak. 
%%\bs

\begin{figure}
\centerline{\protect\hbox{\psfig{file=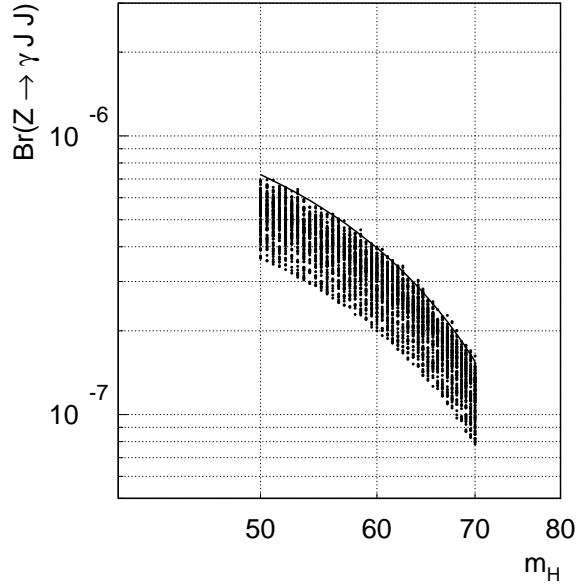,height=8cm}}}
\caption{Width for $Z\ra \gamma  J J$ as a function of $M_h$ for 
our model (points). The standard model result for $Z\ra H \gamma$
(solid line) is shown for comparison.}
\label{babu4_10070}
\end{figure}

\noi
The resulting $Z\ra \gamma J J$ decay branching ratio is shown
in \fig{babu4_10070}. 
Comparing with the results of \fig{babu1_10070} we see that
the strength of this process is essentially the same as that of 
$Z\ra \gamma H$. This can be easily understood. If we change variables to
\beq
x= \frac{2\, M_Z}{M_H\, \Gamma_H}\ E_{\gamma}
\eeq
one can, after some simple algebra, write the total width in the form
\beq
\Gamma(Z\ra \gamma J J)=\int_0^{x_{max}}
\Gamma(Z\ra H \gamma)\ BR(H\ra JJ)\ \frac{1}{\pi}\
\frac{1}{(x-x_0)^2+1}
\eeq
where
\beq
x_{max}={\frac{M_Z^2}{M_H \Gamma_H}} 
\eeq
and
\beq
x_0=\frac{2\, M_Z}{M_H\, \Gamma_H}\ \frac{M^2_Z -M^2_H}{2 M_Z}
\eeq
is the value for which $Q^2=m^2_H$ in terms of the $x$ variable. Now
we notice that
\beq
\int_{-\infty}^{+\infty} \frac{1}{\pi}\
\frac{1}{(x-x_0)^2+1}=1
\eeq
and if the width is very small we can safely set
\beq
\frac{1}{\pi}\ \frac{1}{(x-x_0)^2+1} \simeq \delta(x-x_0)
\eeq
and therefore
\beq
\Gamma(Z\ra \gamma J J) \simeq 
\Gamma(Z\ra H \gamma)\ BR(H\ra JJ)
\eeq

\begin{figure}
\centerline{\protect\hbox{\psfig{file=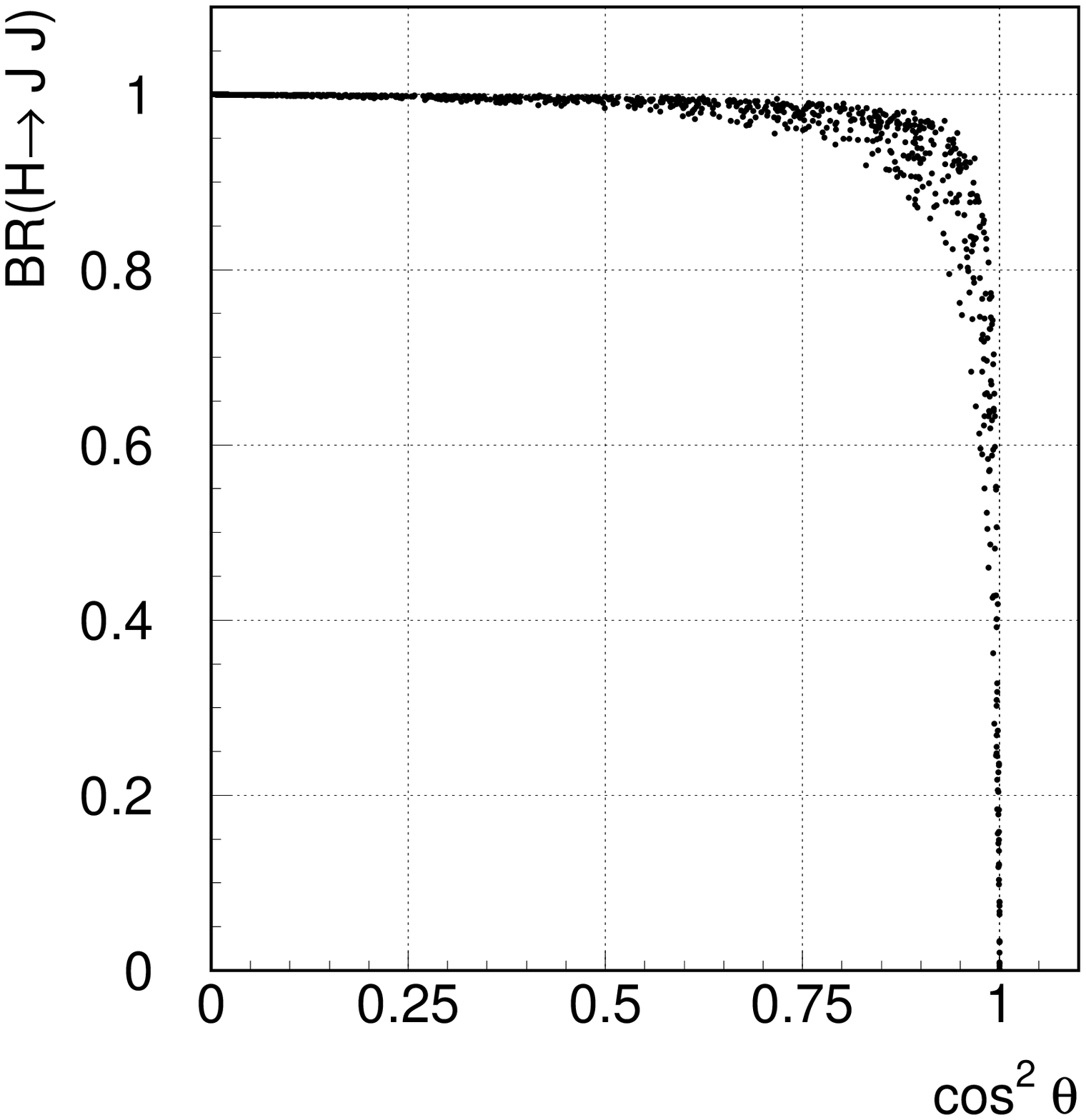,height=8cm}}}
\caption{$BR(H\ra J J)$ as a function of $\cos^2 \theta$.}
\label{babu6}
\end{figure}

\noi
One can see from \fig{babu6} that the $Br(H\ra JJ)$ is very close
to 1 except for the mixing angle in the vicinity of zero as can be
understood from \Eq{hcoup}.

\section{Discussion}

The search for  single-photon plus missing energy events
constitutes one of the classic  experiments in $e^{+} e^{-}$
annihilation. Of course, such events are expected to occur through 
initial state bremsstrahlung with the $Z$ decaying to a $\nu \bar{\nu}$ pair. 
Recently the OPAL collaboration has published a high statistics single-photon 
spectrum that shows some excess of high energy photons above the 
expectations from initial state radiation \cite{opal}. This signal could 
be an interesting hint for physics beyond the standard model. 

In this letter we have studied the rates for single-photon processes
 such as $Z \ra \gamma H$, $Z \ra \gamma J$ and $Z \ra \gamma J J$
where $H$ is a massive CP-even Higgs boson, and $J$ denotes the massless
(or nearly so) CP-odd Majoron associated to the spontaneous violation 
of lepton number around the weak scale.
For this purpose we considered the simple model proposed
in ref. \cite{ewbaryo}. We have demonstrated that in this
simple model the $Z \ra \gamma H$ and $Z \ra \gamma J J$ decays 
may occur with branching ratios compatible with LEP sensitivities. 
That such indirect signals of models of neutrino mass can be sizeable 
is quite remarkable, taking into account that the corresponding neutrino 
masses themselves are very small, as required in order to explain the 
solar neutrino problem. In the model in question the smallness of the 
neutrino masses follows naturally from the fact that they arise only 
at the two-loop level.

The $\gamma$ spectrum associated to these decays  is shown in 
\fig{babu3} and \fig{babu3a}. It is characterized by a spike located 
at a photon energy $E_{\gamma}=(M_Z^2-M_H^2)/(2M_Z)$, determined 
by the possible values of the scalar Higgs boson masses $M_H$. 
The constraints on $M_H$ that follow from the LEP100 experiments
differ from those obtained in the standard model since (i) the 
CP-even Higgs boson neutral-current couplings are somewhat suppressed due
to the admixture of the singlet required to implement the spontaneous 
violation of lepton number and (ii) these CP-even Higgs bosons can decay 
with substantial rates into the invisible channel $JJ$  \cite{inv}. 
Here we showed explicitly how the invisible Higgs decay can be important 
also in conjunction with radiative $Z \ra \gamma H$ decays, leading to a 
sizeable rate for the $Z \ra \gamma E\!\!\!/_T$ signal on the Z peak. 

While LEP200 will play an important role in searching for 
invisibly-decaying Higgs bosons \cite{ebolep2}, high statistics
studies of the single-photon energy spectrum at the Z-pole may 
still be an interesting physics goal, as illustrated through
the model described in this paper.

\section*{Acknowledgements}
 
This work has been supported by DGICYT under Grants PB95-1077
and SAB95-0175 (S.D.R.), by the TMR network grant ERBFMRXCT960090 
of the European Union, and by an Acci\'on Integrada Hispano-Portuguesa.
One of use (S.D.R.) thanks Borut Bajc for discussions and correspondence. 

\newpage

\section*{Appendix}

We will give here the explicit expressions for the various amplitudes
referred in the text. For $A_{SM}$, $A_h$ and $A_k$ we have\cite{BPR}
\beq
A_{SM}=A_W + A_F
\eeq
where
\beqa
A_W&=&4 \cos \theta_W \left[\vbox to 18pt{}
(3 -\tan \theta_W^2)\ J_1(M_Z,M_H,M_W) \right. \cr
&& \cr
&& \left. + \left(-5 +\tan_W^2 \theta_W - \half 
\frac{M_H^2}{M_W^2}(1-\tan \theta_W^2) \right)\ J_2(M_Z,M_H,M_W)
\right]
\eeqa
and
\beq
A_F=\sum_f \frac{4 g^f_V Q_f}{\cos \theta_W}\ \left[\vbox to 14pt{}
-J_1(M_Z,M_H,M_f)+4 J_2(M_Z,M_H,M_f) \right] \ .
\eeq
In the previous equations we have introduced the functions $J_1$ and
$J_2$ defined by
\beqa
J_1(M_Z,M_H,M_W)&=&-  M_W^2\ C_0(M_Z^2,0,M_H^2,M_W^2,M_W^2,M_W^2) \cr
&&\cr
J_2(M_Z,M_H,M_W)&=&
\half \frac{M_W^2}{M_Z^2-M_H^2} \left[ \vbox to 18pt{}1 + 2 M_W^2\
C_0(M_Z^2,0,M_H^2,M_W^2,M_W^2,M_W^2) \right. \cr
&& \cr
&&\left. + \frac{M_Z^2}{M_Z^2-M_H^2}
\left( B_0(M_Z^2,M_W^2,M_W^2)- B_0(M_H^2,M_W^2,M_W^2)\right)\right]
\eeqa
where $B_0$ and $C_0$ are the Passarino-Veltman functions\cite{thvelt}.

%%\bs

\noi
The amplitudes $A_h$ and $A_k$ are given by
\beqa
A_h&=&\frac{4 \sin^2 \theta_W}{\cos \theta_W}\
\left( \frac{\lambda_4\ v^2\ P_{i1} + \lambda_8\ w v\
P_{i2}}{M^2_{h^{\pm}}}\right)\ J_2(M_Z,M_H,M_{h^{\pm}})\cr
&&\cr
A_k&=&\frac{16 \sin^2 \theta_W}{\cos \theta_W}\
\left( \frac{\lambda_5\ v^2\ P_{i1} + \lambda_9\ w v\
P_{i2}}{M^2_{k^{\pm\pm}}}\right)\ J_2(M_Z,M_H,M_{k^{\pm\pm}})
\eeqa

%%\bs

\noi The amplitudes $\hat{A}_{\phi}$, $\hat{A}_h$ and $\hat{A}_k$
coming from the second set of diagrams with quartic vertices are:

\beqa
\hat{A}_{\phi}&=&-4\, \cos \theta_W\, (1- \tan^2 \theta_W)\,
\frac{1}{M_W}\ \frac{\lambda_7}{g}\ J_2(M_Z,M_{JJ},M_W)\cr
&&\cr
\hat{A}_h&=& \frac{4\, \sin^2 \theta_W}{\cos \theta_W}\,
\frac{v\, \lambda_8}{M^2_{h^{\pm}}}\, J_2(M_Z,M_{JJ},M_{h^{\pm}})\cr
&&\cr
\hat{A}_k&=& \frac{16\, \sin^2 \theta_W}{\cos \theta_W}\,
\frac{v\, \lambda_9}{M^2_{k^{\pm\pm}}}\, J_2(M_Z,M_{JJ},M_{k^{\pm\pm}})
\eeqa
where $M^2_{JJ}=Q^2=(q_2+q_3)^2=M_Z(M_Z-2E_{\gamma})$.

\end{document}